\documentclass{aa} 
\usepackage{graphicx}
\begin{document}

\title{On the integrability of stellar motion in
    an accelerated logarithmic potential}

\author{Fathi Namouni\inst{1} \and Massimiliano Guzzo\inst{2} \and Elena Lega\inst{1}}

\institute{Universit\'e de Nice, CNRS, Observatoire de la
 C\^ote d'Azur, BP 4229, 06304 Nice, France \and
 Dipartimento di Matematica Pura ed Applicata,
 Universit\`a degli Studi di Padova,
 via Trieste 63, 35121 Padova, Italy}

\date{Received /Accepted}

\abstract{
An accelerated logarithmic potential models the mean motion of stars
in a flat rotation curve galaxy that sustains a wind system. For stars
outside the galactic wind launching region, the asymmetric removal of
linear momentum by the wind is seen as a perturbing acceleration
superimposed onto the galactic potential.}{We study the integrability of stellar motion in an accelerated logarithmic potential.}{We use surfaces of section of the dynamical system to probe the integrability of motion.}
{We provide numerical
evidence that motion in an accelerated logarithmic potential is
non-integrable.}{Large scale chaotic diffusion occurs {{  for lower values of the projected angular momentum along the direction of acceleration and persists at all values of the angular momentum}} in the outer
part of the galaxy inside the truncation radius where the galactic
acceleration balances the wind-induced acceleration.}


\titlerunning{Stellar motion in an accelerated logarithmic potential}
\authorrunning{Namouni, Guzzo, Lega}

\maketitle

\section{Introduction}
The advent of new observations of galactic winds have shown that these
processes are ubiquitous features in starburst and active galaxies
(Veillet et al. 2005).  In this paper, we lay down the dynamical premise for
studying the effect of asymmetric linear momentum removal from a
galaxy with a flat rotation curve.  Our investigation is motivated by
the recent studies of the dynamical influence of asymmetric stellar
jets on protoplanetary disks and the forming planets they contain 
(Namouni 2005, 2007; Namouni and Zhou 2006; Namouni and
Guzzo 2007). These works have shown that the jet-induced
acceleration leads to a number of dynamical consequences such as
strongly modifying protoplanetary disk structure profiles, forcing the
radial contraction and expansion of the disk, and enhancing the
internal heating of the disk as it follows its varying state of least
energy induced by the jets' variablity. Jet-induced acceleration also
helps truncate the protoplanetary disk, excites the
eccentricity of forming planets, and induces radial migration in
binary systems. In view of the variety of these consequences, we
expect that the wind-induced acceleration alter star orbits and affect the collective response of the
galactic disk.

Modeling the motion of a star in a flat rotation curve galaxy as that
of a test particle moving in a central logarithmic potential, $\log
r$, where $r$ is the distance to the galactic center (Binney and Tremaine 1987), the
asymmetric removal of linear momentum from the galaxy can in turn be
modeled by an additional acceleration that does not depend on the
position and velocity of the star. This approximation is valid as long
as we consider the motion of star outside the wind launching
region. In effect, as such a star revolves around the matter inside
its orbit, it is only affected by the variation of mass and momentum
losses inside its orbit.  It is worth noting that the effect under
consideration is not that of the direct action of the galactic wind on
the star such as that of the wind blowing on the star --which is
physically negligible. Rather it is the indirect effect of the loss of
linear momentum by the galactic disk on the motion of stars outside
the wind launching region.

To set up the dynamical premise of this process, we first
discuss the zero-velocity curves of the accelerated logarithmic
potential as well as its orbits of least energy (section 2). The
surfaces of section allow us to show that orbital motion in the
accelerated logarithmic potential is not integrable (section 3) in
contrast to the accelerated Kepler potential (Epstein 1916, Sommerfeld 1929, Landau and Lifschitz 1969) that
models orbital motion in the presence of a stellar jet (Namouni and Guzzo
2007).
Section~4 contains concluding remarks.

\section{Zero velocity curves and least energy orbits}
Galaxies with asymmetric wind systems lose linear momentum in the
region inside the wind launching radius. The net momentum loss
accelerates the wind launching region with respect to the matter
outside it in the direction opposite to momentum loss
(Namouni 2005, 2007; Namouni and Guzzo 2007). In essence, asymmetric momentum loss from a wind
system accelerates the inner part of the galaxy like a rocket with
respect to its outer part.  Modeling the gravity of a flat rotation
curve galaxy as a logarithmic potential (Binney and
Tremaine 1987), the motion of a
star outside the wind launching region follows the equations:
\begin{equation}
\frac{{\rm d}^2 {\bf x}}{{\rm d} t^2}=-\frac{v_c^2}{|{\bf x}|^2}\,{\bf
  x} +{\bf A}, \label{motion}
\end{equation}
{  where ${\bf x}$ denotes the position of the star in the galaxy}, $v_c$ is the galactic circular velocity and ${\bf A}$ is the
wind-induced acceleration. We examine the effect of momentum loss during a
stationary wind episode and therefore take ${\bf A}$ to be constant in time.
The presence of the acceleration ${\bf A}$ introduces natural units for time
and distance. These are given as:
\begin{equation}
L=\frac{v_c^2}{A} \ \ \ \mbox{and}\ \ \ T=\frac{v_c}{A}. \label{scale}
\end{equation}

\begin{figure}
\begin{center}
\includegraphics[width=0.35\textwidth]{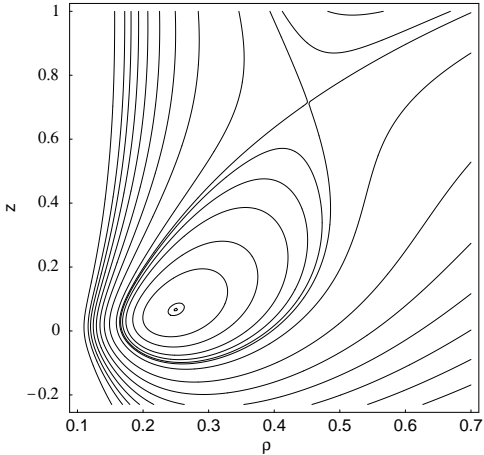}
\caption{Zero velocity curves for the value of $h_z^2=0.058313294$.}
\end{center}
\label{fig1}
\end{figure}

\begin{figure}
\begin{center}
\includegraphics[width=.35\textwidth]{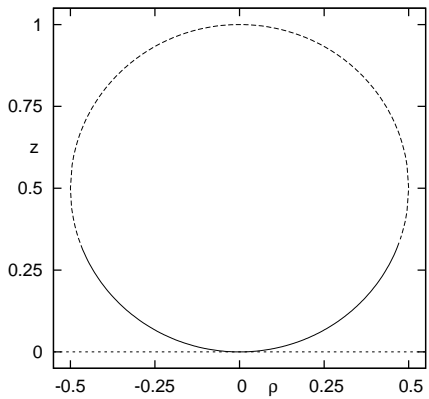}
\caption{Sphere of least energy orbits. The solid (dashed) line denotes stable
  (unstable) orbits. }\label{fig2}
\end{center}
\end{figure}

The former is related to the boundary of the logarithmic potential set by the
equality of the gravitational acceleration and the wind-induced acceleration.
The latter is the typical excitation time for stellar orbits in the presence
of acceleration. In terms of these units, the equations of motion are given
as:
\begin{equation}
\frac{{\rm d}^2 {\bf x}}{{\rm d} t^2}=-\frac{{\bf
  x}}{|{\bf x}|^2}\, +{\bf e}_z, \label{motion2}
\end{equation}
where ${\bf e}_z$ is the unit vector along the direction of acceleration ${\bf
  A}$ that we choose as the $z$--axis of the inertial reference frame. Note
that it is still possible to keep track of the relative strengths of the
logarithmic potential and the constant acceleration. For instance the
logarithmic potential is dominant for $|{\bf x}|\ll1$, and a unit orbital
period corresponds to motion near the boundary where the gravitational
acceleration balances ${\bf A}$.

As the acceleration is constant, the corresponding force derives from the
potential $R=z$ and hence the dynamical system (\ref{motion2}) admits the
constant of motion:
\begin{equation}
E=\frac{1}{2}\, \left(\dot\rho^2 + \rho^2\dot\theta^2+ \dot
  z^2\right)+\frac{1}{2}\, \log \left(\rho^2+z^2\right)-z, \label{energy}
\end{equation}
where $\rho, \theta$ and $z$ are the usual cylindrical coordinates. In
addition to the energy $E$, the projection of the angular momentum ${\bf
  h}={\bf x}\times{\bf v}$ along the direction of acceleration (or
equivalently the $z$--axis), $h_z={\bf h}\cdot {\bf e}_z=\rho^2\dot\theta$, is
a constant of motion. Substituting the expression of $\dot \theta$ in terms of
$h_z$ into Equation (\ref{energy}), the energy equation now reads:
\begin{equation}
E=\frac{1}{2}\, \left(\dot\rho^2 + \dot
  z^2\right)+ \frac{h_z^2}{2\rho^2}+\frac{1}{2}\, \log \left(\rho^2+z^2\right)-z. \label{energy2}
\end{equation}
The zero velocity curves of the system (\ref{motion2}) are obtained from the
effective potential of the energy equation {  -- the last three terms of} (\ref{energy2}). In Figure (1), we
show such curves for the value of $h_z^2=0.058313294$. The region of bounded
motion is restricted in space inside a radius of $\rho^2+z^2\simeq 1$ where
the gravitational attraction exceeds the amplitude of the perturbing
acceleration.  There are two equilibrium (least energy) points located at
($\rho=0.25,\ z=0.067$) and ($\rho=0.452, \ z=0.714$). The former is a stable
orbit and the latter is unstable.

\begin{figure*}
\centering
\hspace{.15cm} \includegraphics[width=.35\textwidth]{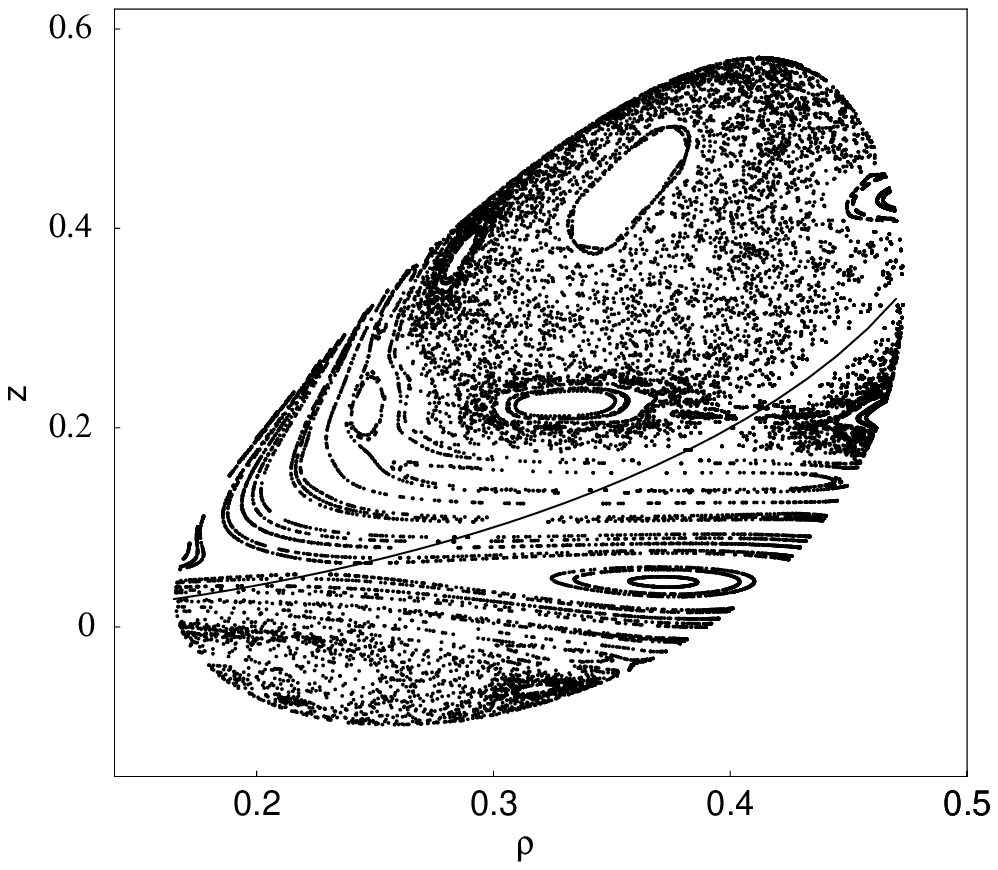}\hspace{-0.21cm} \includegraphics[width=.365\textwidth]{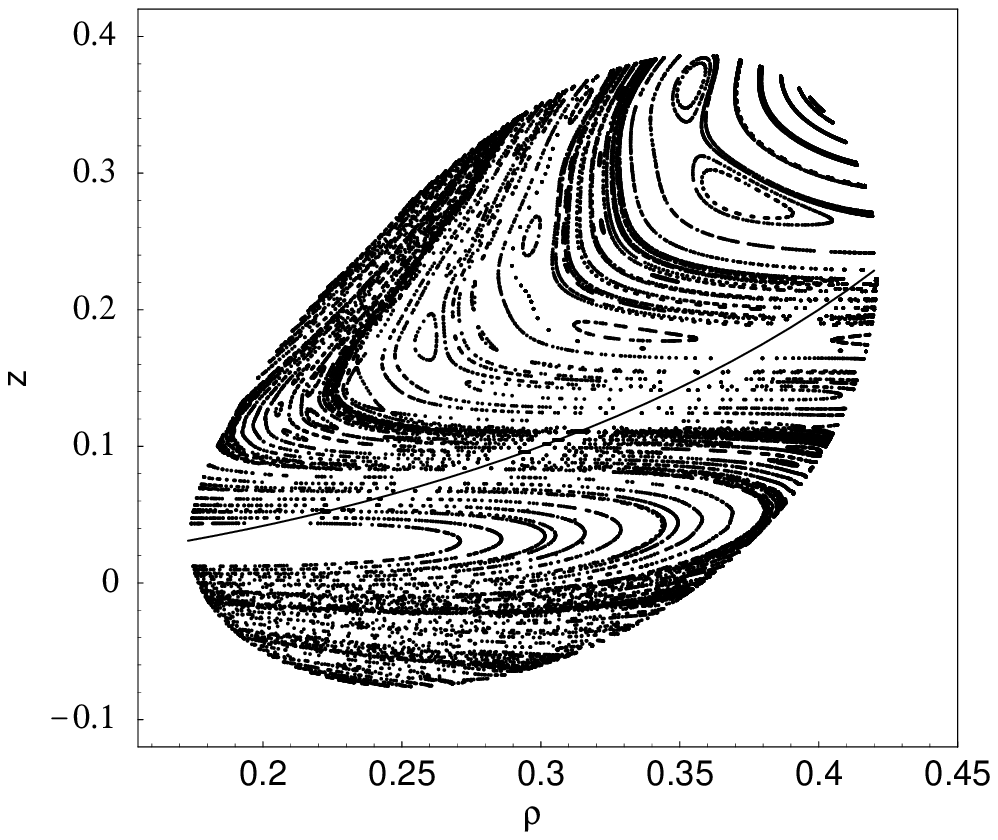}\\
\includegraphics[width=.35\textwidth]{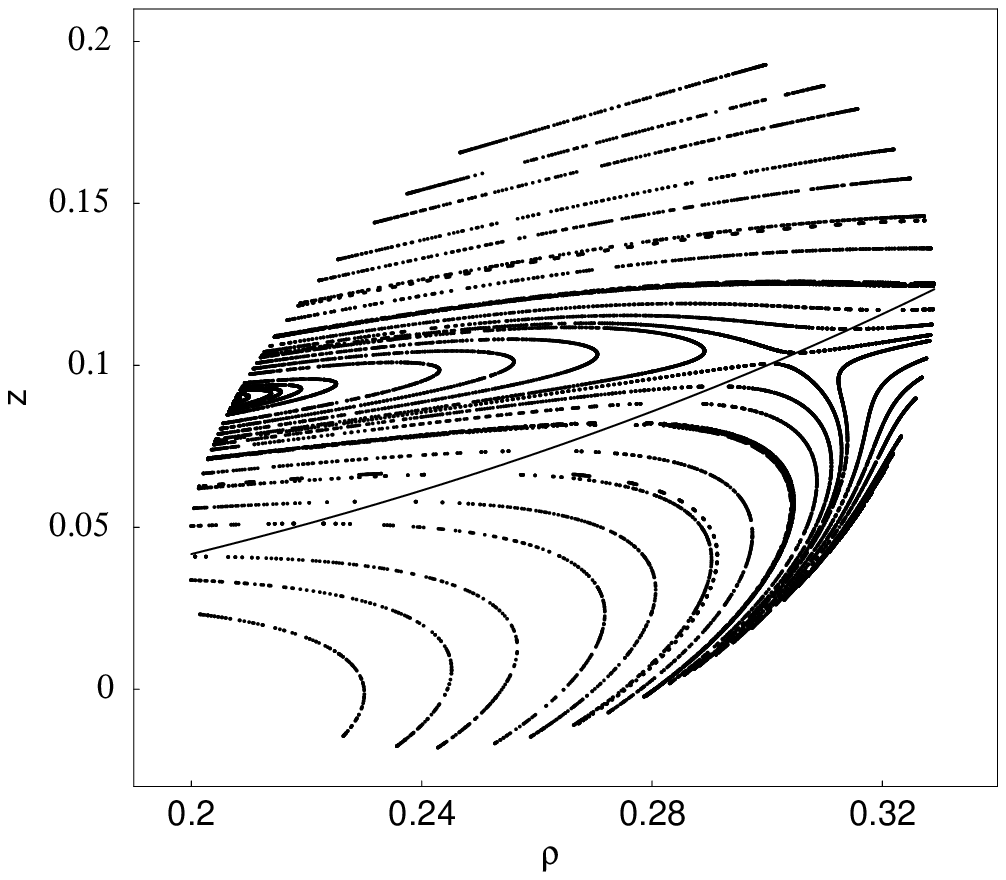}\includegraphics[width=.35\textwidth]{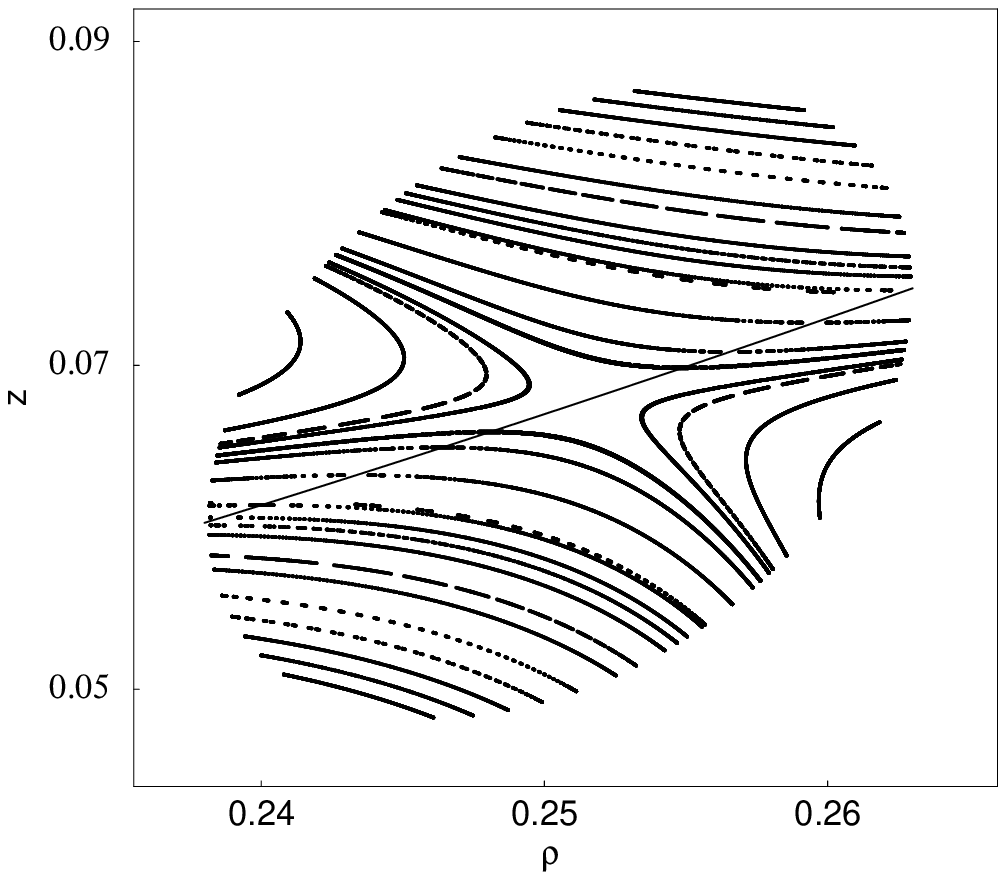}
\caption{Surfaces of section for $h_z^2=0.058313294$. The energy levels are $E=-0.75$ (top left panel), $E=-0.80$ (top right), $E=-0.90$ (bottom left) and $E=-0.95$ (bottom right). The solid
  line is the sombrero profile 
  $z=(1-\sqrt{1-4\rho^2})/2$ that divides space into $\ddot z <0$ orbits
  (above the line) and $\ddot z >0$ orbits
  (below the line).}
\end{figure*}

\begin{figure*}
\centering
\hspace{-.1cm} \includegraphics[width=.35\textwidth]{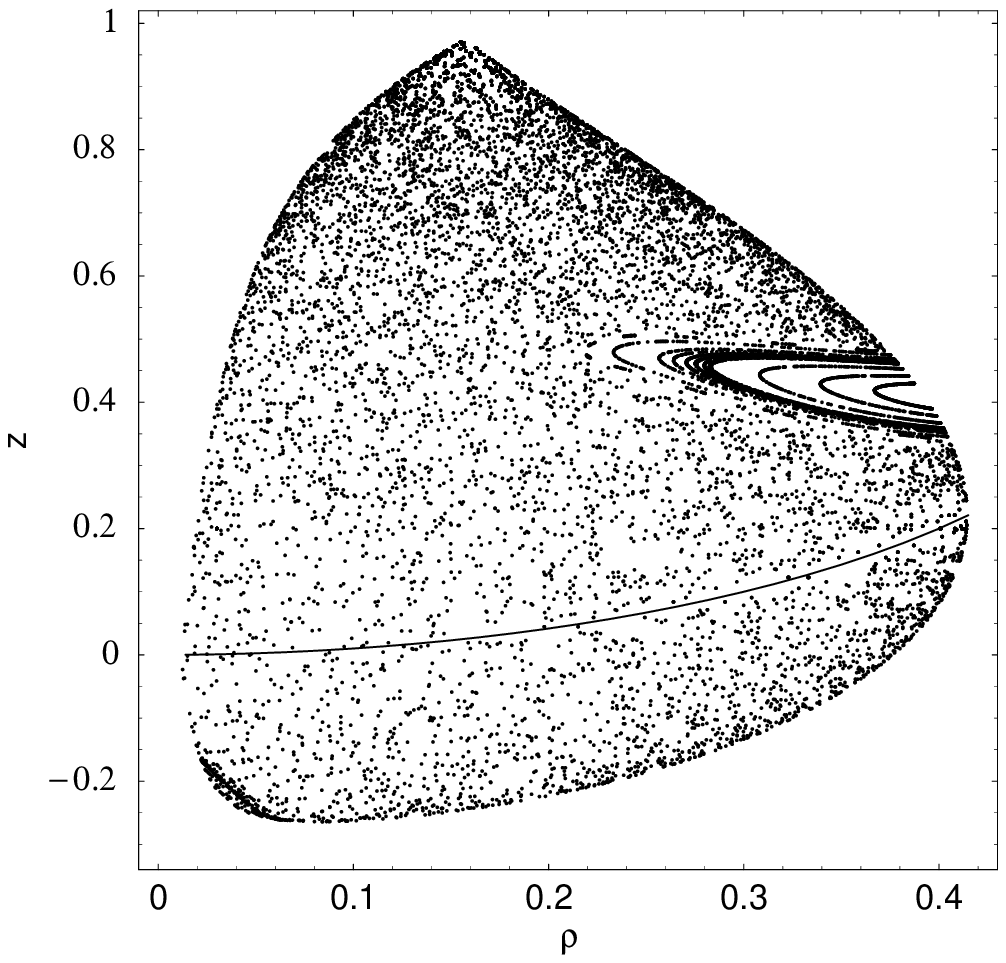}\hspace{-0.cm} \includegraphics[width=.343\textwidth]{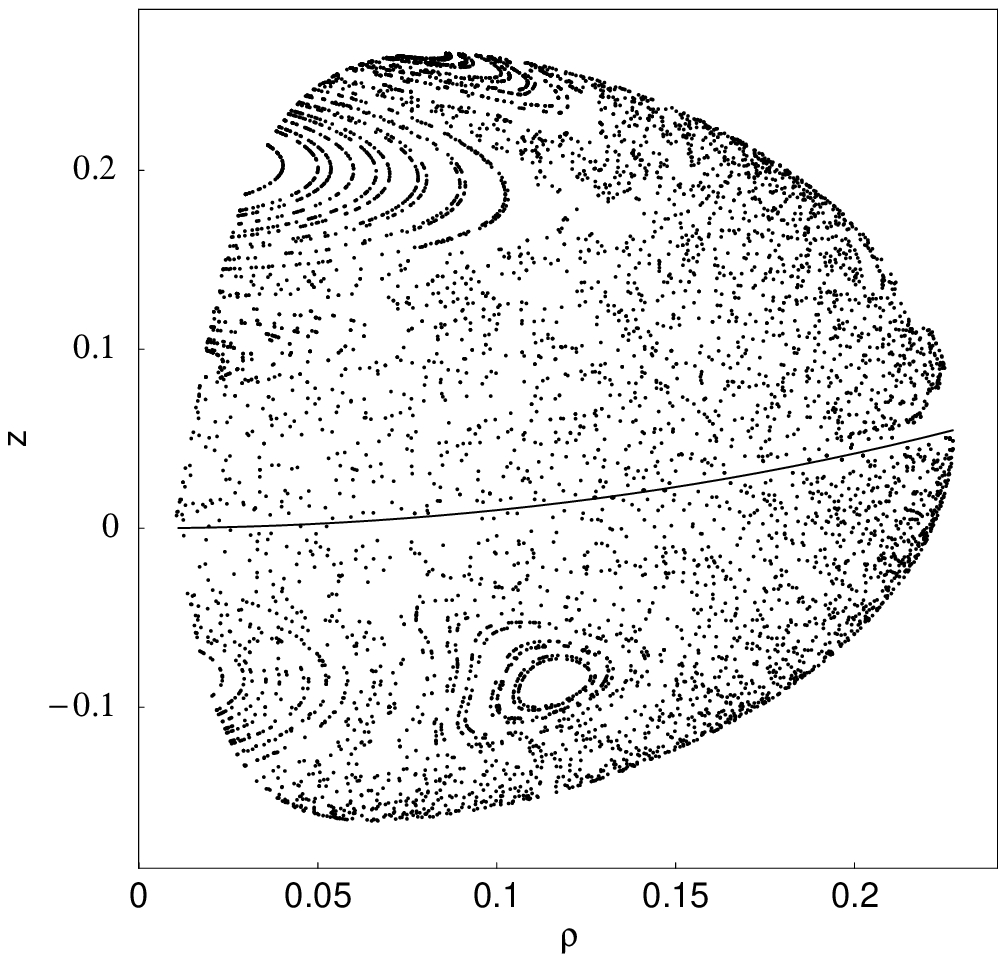}\\
\includegraphics[width=.35\textwidth]{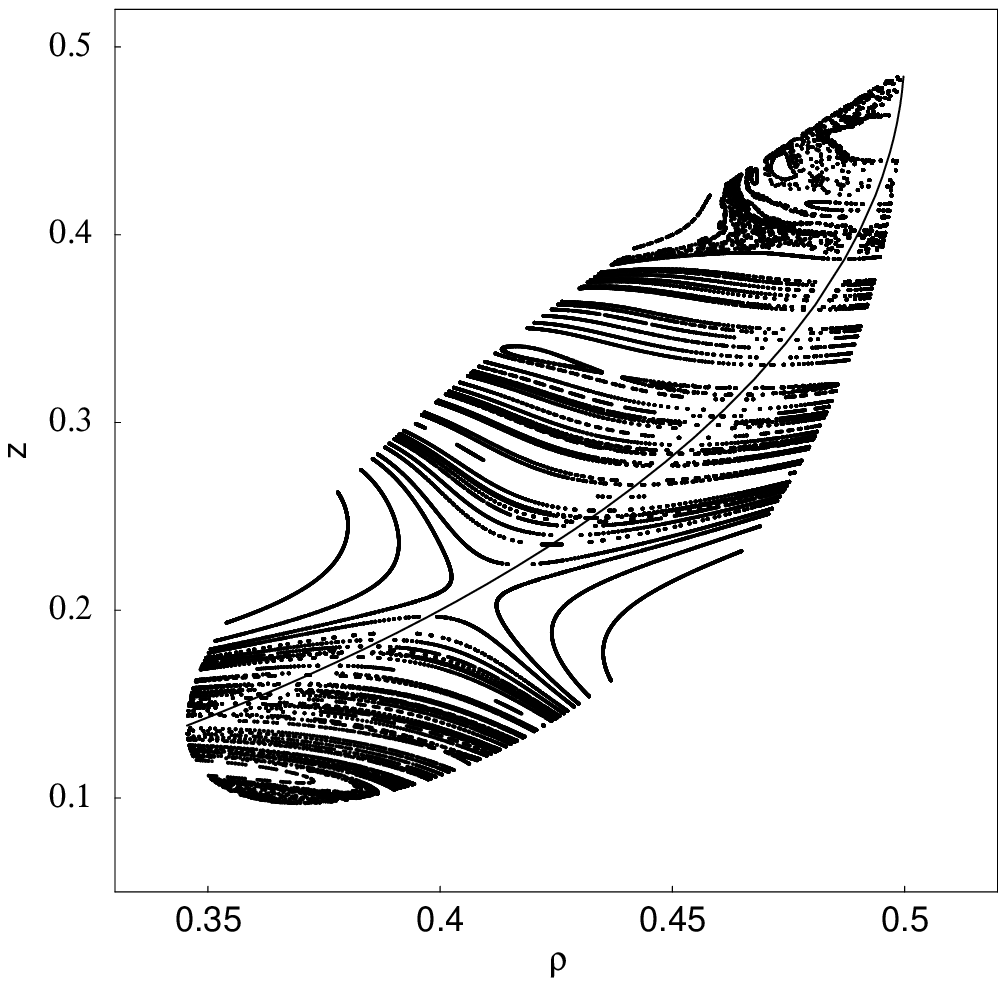}\includegraphics[width=.35\textwidth]{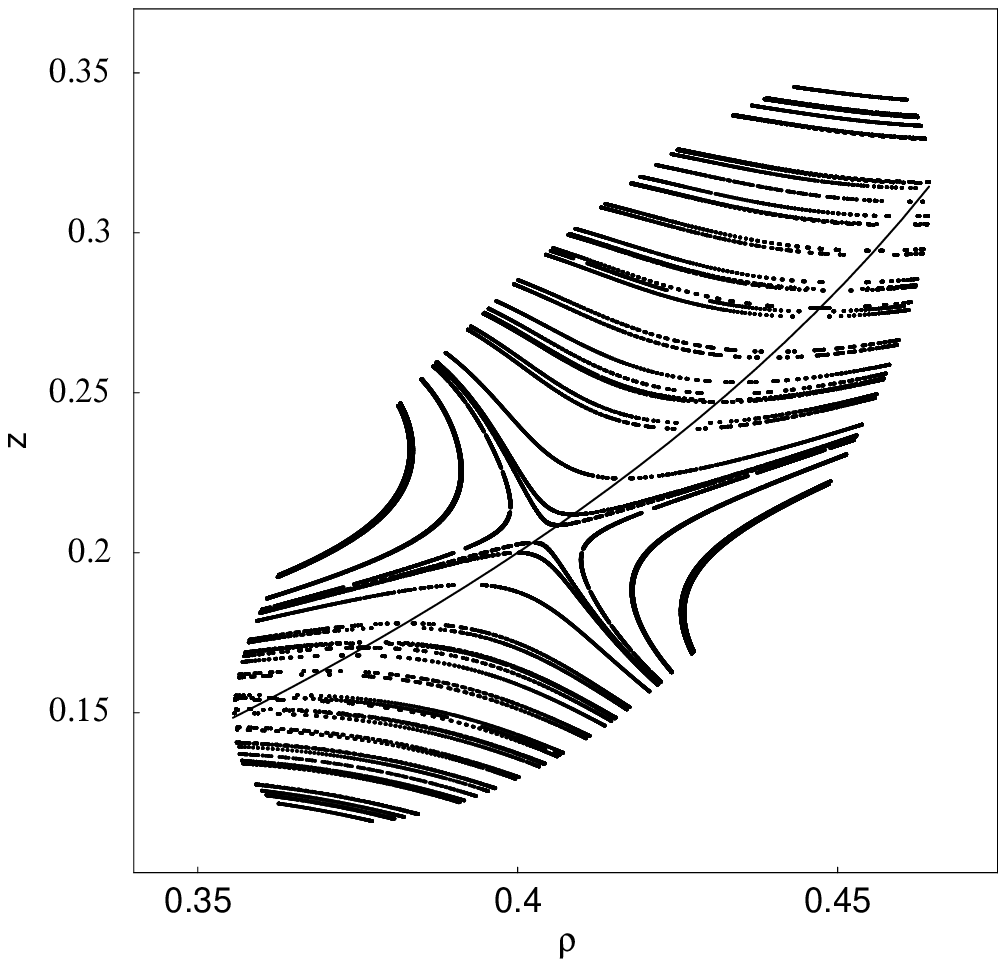}
\caption{  Surfaces of section for different values $h_z^2=6.24\times 10^{-4}$ (top row) and $h_z^2 =0.128$ (bottom row). For   $h_z^2=6.24\times 10^{-4}$, the energy levels are $E=-0.97485$ (top left) and $E=-1.5$ (top right).  For $h_z^2=0.128$, the energy levels are $E=-0.590575$ (bottom left) and $E=-0.596$ (bottom right).  The solid 
  line is the sombrero profile  $z=(1-\sqrt{1-4\rho^2})/2.$  }
\end{figure*}

{  We determine the properties of least energy orbits by equating with zero the partial
derivatives of $E$ with respect to $\rho,\ z, \dot\rho$ and $\dot z$.  The
derivatives with respect to $\dot\rho$ and $\dot z$ show that least energy
orbits are circular and planar. The derivative with respect to $\rho$ gives
the expression of $h_z$:
\begin{equation}
h_z^2=\frac{\rho^4}{\rho^2+z^2}\label{hz2}
\end{equation}
or equivalently the orbital rotation rate
$\Omega=\dot\theta=h_z\rho^{-2}$ in terms of $\rho$ and $z$. Lastly, the derivative with
respect to $z$, which is the vertical projection of the equation of motion, shows that least energy orbits hover above the center of
gravitational attraction according to the equations:
\begin{equation}
z=\frac{1\pm\sqrt{1-4\rho^2}}{2}. \label{leoz}
\end{equation}
From this equation, it is seen that the locus of least energy circular orbits is a
sphere centered on ($\rho=0$, $z=1/2$) of radius $1/2$ (Figure 2).  The upper and lower signs do not correspond to the stability of the
equilibrium orbits. To determine which orbits are stable on the least energy sphere, we first note that every least energy orbit shown in Figure (2) has a different value of the projected angular momentum, $h_z$, and the energy, $E$. From the zero velocity curves (Figure 1), we know that for a given value of $h_z$ there are two least energy orbits, one stable and one unstable --- the smallest value of $h_z$, zero, must corresponds to the two least energy points at $\rho=0$ and $z=0$ (stable) and  $z=1$ (unstable).  Using Equation (\ref{leoz}), it is possible to express  $h_z$ (\ref{hz2}) as a function of $z$ for least energy orbits and solve for the
values of $z$ (and hence $\rho$) that determine the stable and unstable orbits.} This substitution yields:
\begin{equation}
h_z^2=z\,(z-1)^2.
\end{equation}
This shows that $h_z^2$ has a maximum at:
\begin{equation}
z_{\rm b}=1/3, \ \ \ \mbox{and}   \ \ \ \rho_{\rm b}=\sqrt{2}/3,
\end{equation}
and as expected $h_z$ vanishes for $z=0$ and $z=1$ ($\rho=0$).  
The maximum value of $h_z^2$ is 4/27 and corresponds to the boundary
of stable least energy orbits. It also defines the truncation radius of the galaxy. The condition
\begin{equation}
|h_z|< 2/3\sqrt{3}
\end{equation}
is therefore necessary for bounded motion to occur.  The profile of
stable and unstable orbits is shown in Figure (2) {  with solid and dashed lines respectively}. The locus of the
stable orbits in the $\rho z$--plane has a profile similar to (but
less flat at its base than) the sombrero-shaped profile associated
with the accelerated Keplerian potential (Namouni 2007; Namouni and Guzzo 2007). 
For consistency
with the Keplerian case, we call the locus of the stable orbits of an
accelerated galactic potential, the sombrero profile. An additional
difference between the accelerated Keplerian and logarithmic
potentials is that the former admits a global third integral of motion
implying that orbits are everywhere regular --even in the vicinity of
the zero-velocity separatrix and the unstable least energy orbit.  The
accelerated logarithmic potential does not have a global third integral. This
we show in the next section.

\section{Surfaces of Section}

The integrability of stellar motion in the accelerated logarithmic potential
may be determined numerically by drawing surfaces of section of the dynamical
system (\ref{motion2}).  Sections in phase space are chosen for (i) a constant
energy, $E$, (ii) a constant projected angular momentum, $h_z$, and (iii) with
the condition $\dot z=0$. Sections usually require a crossing direction such
as that given by $\ddot z>0$. However, as we find it convenient to represent
sections in the $\rho z$--plane, we can readily determine the sign of $\ddot
z$ everywhere in that plane.  As $\ddot z$ equals the vertical component of
the gravitational potential augmented by the perturbing acceleration, $\ddot
z$ vanishes exactly on the sphere of least energy (Equation \ref{leoz}). The
sign of $\ddot z$ is negative inside the least energy sphere and positive
outside it. This allows us to explore numerically the whole area inside the
zero velocity curve associated with the values of $E$ and $h_z$, and know the
sign of $\ddot z$. Accordingly, we plot the curve of $\ddot z=0$ (Equation
\ref{leoz}) on all surfaces of section which are now allowed to include both
$\ddot z>0$ orbits and $\ddot z<0$ orbits.

Using the same value of the sombrero orbit as that of Figure (1), $\rho = 0.25$,
we set $h_z^2=0.058313294$ and compute the surfaces of section for $E=-0.95, \ 
-0.90,\ -0.80$ and $-0.75$. The value $E=-0.95$ corresponds to a
small region around the sombrero orbit. Increasing the value of $E$ widens the
area of possible motion. Orbits have been computed for a duration $t=200$ as
well as for various values of $\rho$, $z$ and $\dot z$ whereas $\dot \rho$ is
obtained from the energy equation.

The section for $E=-0.75$ (Figure 3, top left) shows that motion in the accelerated
logarithmic potential is not integrable with large chaotic regions developing
for smaller and larger elevations $z$. Irregular motion is milder near the
plane $z=0$ that contains the gravitational center. Noting that the surface of
section is asymmetric with respect to the sombrero profile
$z=(1-\sqrt{1-4\rho^2})/2$, restricting the surface of section to positive
$\ddot z$ only would have lead to a loss of information on the motion inside
the least energy sphere. The lowering of the value of $E$ shrinks down the
region of possible motion and as well as that of irregular motion (Figure 3, top right). In particular for $E=-0.90$  motion is further restricted to a small region
around the sombrero orbit $(\rho=0.25, z=0.067)$ and an unstable point appears
very near it (Figure 3, bottom panels). Lowering the energy further makes the unstable point
converge towards the sombrero orbit ($E=-0.95$, Figure 3, bottom right) implying the
presence of a mild separatrix-type diffusion in the vicinity of the least
energy orbit.

{  The extent of the regions of chaotic motion also depends on the value of the angular momentum $h_z$. We find that the smaller the projected angular momentum $h_z$ the larger the chaotic motion regions. This is shown in Figure (4) where we plot surfaces of section for $h_z^2=6.24\times 10^{-4}$ and $0.128$. The former corresponds to a stable sombrero orbit at $\rho=0.025$ and $z=6.25\times 10^{-4}$ and the latter to $\rho=0.4$ and $z=0.2$. On Figure (1), these two orbits are on each side of that corresponding to Figure (3). For   $h_z^2=6.24\times 10^{-4}$, the energy levels are $E=-0.97485$ (top left) and $E=-1.5$ (top right). The former has been chosen close to the energy value of the unstable least energy orbit. The region of chaotic motion is large and remains so even when the energy value is lowered ($E=-1.5$) so that the zero velocity curve extent (i.e. the size of the surface of section) is 4 times smaller in the vertical direction that it is for  $E=-0.97485$. For $h_z^2=0.128$, the energy levels are $E=-0.590575$ (bottom left) and $E=-0.596$ (bottom right). As for the previous value of $h_z$, the former energy has been chosen close to the value of the unstable least energy orbit. This time, the region of chaotic motion is confined to the vicinity of the unstable orbit and disappears when the energy value is lowered slightly ($E=-0.596$) so that the zero velocity curve extent is only  1.4 times smaller in the vertical direction that it is for  $E=-0.590575$.  We conclude that for the smaller $h_z$,  the extent of the chaotic region is much larger than that of Figure (3) and irregular motion dominates the surface of section even when the energy value is far from that of the unstable least energy orbit. For the larger value of $h_z$, extended chaotic motion is confined to the vicinity of the unstable least energy orbit as soon as the energy level lowered away from that of the unstable orbit, no region inside the zero velocity curve shows large chaotic motion.   
}

\section{Concluding remarks}

In this paper we showed that stellar motion in an accelerated
logarithmic potential is not integrable unlike motion in an
accelerated Keplerian potential. We have used zero-velocity curves,
least energy orbits as well as surfaces of section to determine the
global topology of orbital motion. {  We find that for a given projected
angular momentum, the larger the area of the zero-velocity curve in
$\rho z$--space, the larger the fractional size of the chaotic
motion region.  The size of the chaotic motion regions also depends on the value of the projected angular momentum $h_z$. The smaller $h_z$, the larger the extent the chaotic motion region. Smaller values of $h_z$ correspond to inner sombrero 
orbit  (i.e. $\rho\ll 1$) while larger values of $h_z$ correspond to zero velocity curves whose corresponding motion is confined near to the truncation radius of the galaxy  (i.e. the stable and unstable least energy orbits become close).

For the diffusion that we have uncovered in this work to affect stellar orbits, the local diffusion time has to be smaller than the age of the galaxy. This implies that in the outer regions where the dynamical time  (i.e. the orbital period $2\pi r/v_c$) is comparable to the age of the galaxy, diffusion does not have enough time to occur. The reason is that the local diffusion time is always and everywhere larger than the local dynamical time. For instance, the scatter of points in the top left panel Figure (3) has been obtained after 200 time units --a time unit being the dynamical time at the truncation radius of the galaxy, $T$ (\ref{scale}). For a typical galaxy, such a timespan is larger than the age of the Universe. 

Diffusion therefore affects stellar orbits in an accelerated flat rotation curve galaxy for low values the projected angular momentum $h_z$ and well inside the truncation radius (i.e. the outer edge of the sombrero profile).} Stellar orbits near sombrero orbits  are more regular as
diffusion is negligible. In these parts of the galaxy, it is possible to make use of secular perturbation techniques to ascertain the
evolution of stellar orbits whose periods are small compared to the duration of the wind episode.

\begin{acknowledgements}
The authors thank the referee for useful comments.
\end{acknowledgements}

\bibliographystyle{aa}

\end{document}